\documentclass[10pt,floatfix]{revtex4}
                                                                                
\usepackage{graphicx}
\usepackage{mathrsfs}
\usepackage{amsmath}
\usepackage{amsfonts}
\usepackage{amssymb}
\usepackage{epsfig}                                                                                
\usepackage{theorem}
\theorembodyfont{\upshape}
\newtheorem{conjecture}{Conjecture}

\newcommand{\be}{\begin{equation}}
\newcommand{\ee}{\end{equation}}
\newcommand{\bea}{\begin{eqnarray}}
\newcommand{\eea}{\end{eqnarray}}

\input epsf.sty

\newlength{\diaght}
\newlength{\diaghtwo}
\newlength{\diaghtthree}
\newlength{\diagshift}
\newcommand{\cs}{\raisebox{\diagshift}{\includegraphics[height=\diaght]{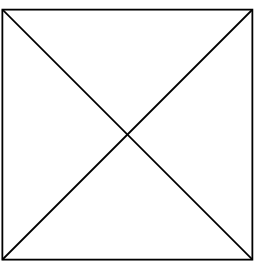}}}
\newcommand{\hs}{\raisebox{\diagshift}{\includegraphics[height=\diaght]{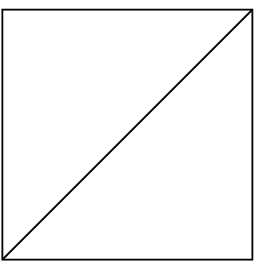}}}
\newcommand{\sq}{\raisebox{\diagshift}{\includegraphics[height=\diaght]{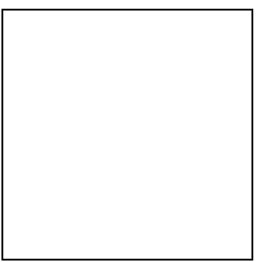}}}
\newcommand{\cso}{\raisebox{\diagshift}{\includegraphics[height=\diaght]{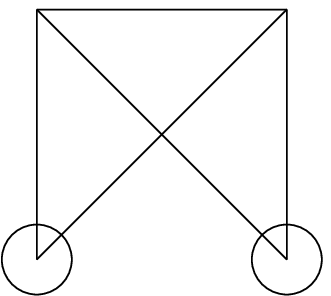}}}
\newcommand{\gsqo}{\raisebox{\diagshift}{\includegraphics[height=\diaght]{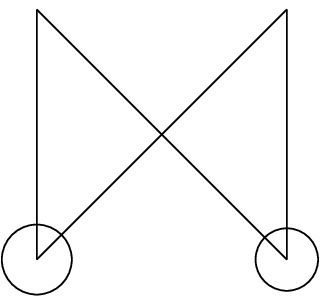}}}
\newcommand{\hhso}{\raisebox{\diagshift}{\includegraphics[height=\diaght]{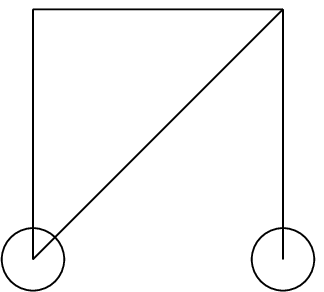}}}
\newcommand{\sqo}{\raisebox{\diagshift}{\includegraphics[height=\diaght]{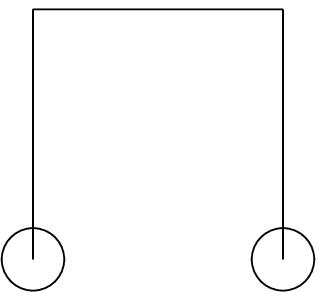}}}
\newcommand{\btwo}{\raisebox{\diagshift}{\includegraphics[height=\diaghtwo]{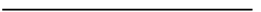}}}
\newcommand{\bthree}{\raisebox{\diagshift}{\includegraphics[height=\diaght]{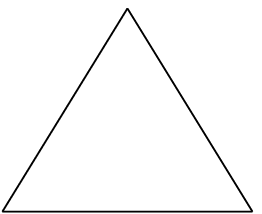}}}
\newcommand{\rhring}{\raisebox{\diagshift}{\includegraphics[height=\diaghtthree]{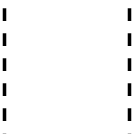}}}

\begin{document}

                                                                                
\title{The fourth virial coefficient of a fluid of hard spheres in odd dimensions}

\author{I. Lyberg
\footnote{e-mail ilyberg@grad.physics.sunysb.edu}}
\affiliation{ Institute for Theoretical Physics, State University of New York,
 Stony Brook,  NY 11794-3840}
\date{\today}
\preprint{YITPSB-04-50}

\begin{abstract}
The fourth virial coefficient is calculated exactly for a fluid of hard spheres in odd dimensions up to 11.   
\end{abstract}
 
\maketitle                                                                             
                                                                               
\begin{flushright}
 {\tt YITP-SB-04-50}
\end{flushright}
                                                                                
\section{Introduction}
\label{intro}

The virial expansion of a fluid of hard spheres has been studied for more than a century, but has not yet been solved. Clisby and McCoy \cite{C/M1} recently made some progress by calculating the exact value of the fourth virial coefficient in even dimensions 4 through 12. Previously the fourth virial coefficient had been analytically calculated in three dimensions by Boltzmann \cite{B} in 1899, and in two dimensions by Rowlinson \cite{R1} in 1964. The second virial coefficient in three dimensions had already been calculated by van der Waals \cite{vdW} and the third had been calculated independently by Boltzmann \cite{B3} and J\"{a}ger \cite{Jaeger}. Luban and Baram \cite{L/B} found functions that give the second and third virial coefficients in any dimension. They also found general functions for the two lower order diagrams in the fourth virial coefficient, but they did not find such a function for the complete star. Luban and Michels \cite{LM} later obtained the four point complete star in 4 and 5 dimensions analytically as triple infinite sums.

In this paper we compute exactly the general expression for the fourth virial coefficient of a fluid of hard spheres in odd dimensions up to 11.

We consider the low density expansion of a fluid of $D$-dimensional hard spheres. The position of the $i$th particle is written as $\mathbf{r}_{i}$, and the distance between two points is written as $r_{ij}=|\mathbf{r}_{i}-\mathbf{r}_{j}|$. $\phi(r_{ij})$ is the pair potential of two points. For a fluid of hard spheres this is
\begin{eqnarray}
\phi(r_{ij}) = \left\lbrace \begin{array}{ll}
            \infty  & \mbox{if $r_{ij}<\sigma$} \\
            0 & \mbox{if $r_{ij}\geq\sigma$}
\end{array} \right.
\label{hspot}
\end{eqnarray}
The Hamiltonian is 
\begin{eqnarray}H=\sum_{i<j=2}^{N}\phi(r_{ij})+\sum_{i=1}^{N}\frac{p_{i}^2}{2m}
\end{eqnarray}
For low densities
\begin{eqnarray}P/kT=\rho+\sum_{n=2}^{\infty} B_{n}\rho^n 
\end{eqnarray}
where $\rho=N/V$.

There are several systematic ways to formalise the calculation of the virial coefficients $B_{n}$. One of these methods is the Mayer expansion \cite{MM}, \cite{U/F}. In this expansion a function called the Mayer $f$ function is defined:
\begin{eqnarray}f(r_{ij})=e^{-\phi(r_{ij})/kT}-1
\end{eqnarray}
The virial coefficients are given by \cite{U/F}
\begin{eqnarray}B_{n+1}=-\frac{n}{n+1}\frac{1}{n!V}\int_{V}...\int_{V}V_{n+1}(\mathbf{r}_{1},...,\mathbf{r}_{n})d^D\mathbf{r}_{1}...d^D\mathbf{r}_{n}
\label{bn}
\end{eqnarray}
where $V_{n+1}$ is the collection of labelled biconnected Mayer diagrams with $n$ points. Each bond of these diagrams represents a function $f(r_{ij})$ in the integrand of (\ref{bn}). Explicitly
\begin{eqnarray}  B_2 = -\frac{1}{2}\int_{V}f(r_{12})d^D\mathbf{r}_{2}= -\frac{1}{2} \,\btwo
\label{b2mayereq}
\eea
\bea B_3 = -\frac{1}{3}\int_{V}f(r_{12})f(r_{13})f(r_{23})d^D\mathbf{r}_{2}d^D\mathbf{r}_{3}= -\frac{1}{3} \,\bthree
\label{b3mayereq}
\end{eqnarray}
\begin{eqnarray}  B_4 =-\frac{1}{8} \,\cs-\frac{3}{4} \,
\hs-\frac{3}{8} \, \sq 
\label{b4mayereq1}
\end{eqnarray}

For hard spheres the Mayer $f$ function is
\begin{eqnarray}
f(r_{ij}) = \left\lbrace \begin{array}{ll}
            -1  & \mbox{if $r_{ij}<\sigma$} \\
            0 & \mbox{if $r_{ij}\geq\sigma$}
\end{array} \right.
\end{eqnarray}
From now on we let $\sigma=1$. If a hard sphere potential is considered, $B_{2}$ and $B_{3}$ can be explicitly evaluated using the formulae \cite{L/B}
\begin{eqnarray}B_{2}=\frac{\pi^{D/2}}{2\Gamma(D/2+1)}
\end{eqnarray}
and
\begin{eqnarray}
\frac{B_{3}}{B_{2}^2}=\frac{4\Gamma{(1+D/2)}}{\pi^{1/2}\Gamma{((1+D)/2)}}\int_{0}^{\pi/3}\sin^D{\varphi}~d\varphi
\label{b3formula}
\end{eqnarray}
The integral in (\ref{b3formula}) is easily evaluated. Let $m$ be any positive integer, and let $u$ be any positive number. According to reference \cite{GR}, p. 159 
\bea \nonumber\int_{0}^{u}\sin^{2m}{x}~dx=\frac{(2m-1)!!}{2^mm!}u~~~~~~~~~~~~~~~~~~~~~~~~~~~~~~~~~~~~~~~~~~~~~~~~~~~~~~~~~~~~~~~~~~~~~~~~~~~~~~~~~~~~~~~\\
-\frac{\cos{u}}{2m}\left\{\sin^{2m-1}{u}+\sum_{k=1}^{m-1}\frac{(2m-1)(2m-3)...(2m-2k+1)}{2^k(m-1)(m-2)...(m-k)}\sin^{2m-2k-1}{u}\right\}
\label{inteven}
\eea
and
\bea \nonumber\int_{0}^{u}\sin^{2m+1}{x}~dx=\frac{2^mm!}{(2m+1)(2m-1)!!}~~~~~~~~~~~~~~~~~~~~~~~~~~~~~~~~~~~~~~~~~~~~~~~~~~~~~~~~~~~~~~~~~~~~~~~~~~~~~~~~~~~~~~\\
-\frac{\cos{u}}{2m+1}\left\{\sin^{2m}{u}+\sum_{k=0}^{m-1}\frac{2^{k+1}m(m-1)...(m-k)}{(2m-1)(2m-3)...(2m-2k-1)}\sin^{2m-2k-2}{u}\right\}
\label{intodd}
\eea
Table \ref{b2b3} shows the values of the second and third virial coefficients in dimensions two to eight.
The evaluation of $B_{4}$ using the Mayer formalism is much more difficult. This calculation was done in two dimensions by Rowlinson in 1964 \cite{R1}, and recently in dimensions 4, 6, 8, 10 and 12 by Clisby and McCoy \cite{C/M1}. The results are shown in table \ref{b4table}. Explicitly, the method that was used was to calculate the volume of the intersection of three spheres
\be
v_{D}(r_{12},r_{13},r_{23})=-\int_{V}f(r_{14})f(r_{24})f(r_{34})d^D\mathbf{r}_{4}
\label{vol}
\ee
as an intermediate step. It follows from (\ref{vol}) that
\be
\,\cs=-\int_{V}\int_{V}\int_{V}f(r_{13})f(r_{23})f(r_{24})v_{D}(r_{12},r_{13},r_{23})d^D\mathbf{r}_{1}d^D\mathbf{r}_{2}d^D\mathbf{r}_{3}
\label{row}
\ee
Rowlinson had prevoiusly calculated $v_{3}(r_{12},r_{13},r_{23})$ \cite{R2}, but no one has so far calculated the three dimensional complete star using (\ref{row}). The reason is that there are elliptic integrals in the odd dimensional case that cancel in the even dimensional case.

The history of the computation of $B_{4}$ in 3 dimensions dates back to the end of the nineteenth century \cite{NK}. Van der Waals formulated a sum of integrals which he thought would give $B_{4}$. However, there was one integral which he could not evaluate (This was the one which is today called the complete star). Van Laar managed to evaluate this integral and published his result in 1899 \cite{L}. Boltzmann contested van der Waals' formulation of the problem, and using the correct virial series expansion he published the correct result in the same year \cite{B}. Boltzmann's result was
\begin{eqnarray}\frac{B_{4}}{B_{2}^3}=\frac{2707}{4480}+\frac{219}{2240}\frac{\sqrt{2}}{\pi}-\frac{4131}{4480}\frac{\arccos(1/3)}{\pi}
\end{eqnarray}
This result was confirmed in 1952 by Nijboer and van Hove \cite{N/H} using what is called the two center formalism.

We shall extend this two centre computation to odd dimensions through $D=11$. The analytic results are given in table \ref{b4table}.

The virial coefficient $B_{4}$ in odd dimensions has previously been computed by Monte Carlo methods by Ree and Hoover \cite{ReeHoover0} and Clisby and McCoy \cite{C/M2}. These numerical results gave the first demonstration that the hard sphere virial coefficients can be negative. The question of negativity of hard sphere virial coefficients is of great theoretical importance, and in dimensions $D\leq4$ Monte Carlo investigations have thus far seen only positive $B_{n}$ for $n\leq10$. The reliability of the conclusions relies on the precision of the Monte Carlo evaluation and therefore it is of interest to have an independent confirmation of the error quoted with the Monte Carlo result. Hence it is most interesting to compare the exact result of $B_{4}$ for $D=5,~7,~9,~11$ with the previous Monte Carlo results. This comparison is made in table \ref{b4table}. Here we see that the exact results lie within 3 times the quoted error of reference \cite{C/M2}. This gives some measure of the accuracy of the Monte Carlo results of reference \cite{C/M2}. The exact results for $D=4,~5$ are within the errors of the computation by reference \cite{Bi}. On the other hand, the estimated value of the infinite triple sum for $D=5$ given by reference \cite{LM} is far outside the stated error.

In section II we review the relation between the two center formalism and the Mayer formalism. In section III we use the two center formalism to evaluate $B_{4}$ in dimensions 5, 7, 9 and 11. We conclude in section IV with a discussion of the prospects of obtaining $B_{5}$ and other higher order coefficients exactly.

\begin{table}
\caption{The second and third virial coefficients}

\begin{tabular}{|l|l|l|l|} \hline\hline
$D$ & $B_{2}$ & $B_{3}/B_{2}^2$ & decimal expansion\\ \hline
2 & $\pi/2$ & $4/3-\sqrt{3}/\pi$ & 0.78200...\\
3 & $2\pi/3$ & $5/8$ & 0.625\\
4 & $\pi^2/4$ & $4/3-(3/2)\sqrt{3}/\pi$ & 0.50634...\\
5 & $4\pi^2/15$ & $53/2^7$ & 0.41406...\\ 
6 & $\pi^3/12$ & $4/3-(9/5)\sqrt{3}/\pi$ & 0.34094...\\
7 & $8\pi^3/105$ & $289/2^{10}$ & 0.28222...\\
8 & $\pi^4/48$ & $4/3-(279/140)\sqrt{3}/\pi$ & 0.23461...\\ \hline\hline
\end{tabular}
\label{b2b3}
\end{table}

\begin{table}[h]
\caption{Exact and numerical values of the fourth virial coefficient}

\begin{tabular}{|l|l|l|} \hline\hline
$D$ &
$B_{4}/B_{2}^3$ & decimal expansion\\ \hline
2 & $2-\frac{9}{2}\frac{\sqrt{3}}{\pi}+10\frac{1}{\pi^2}$ & $\;\;\>0.53223180...$\\
  & & \\
3 & $\frac{2707}{4480}+\frac{219}{2240}\frac{\sqrt{2}}{\pi}-\frac{4131}{4480}\frac{\arccos{(1/3)}}{\pi}$ & $\;\;\>0.28694950598...$\\
  & & \\
4 & $2-\frac{27}{4}\frac{\sqrt{3}}{\pi}+\frac{832}{45}\frac{1}{\pi^2}$ & $\;\;\>0.15184606235...$\\
  & & $\;\;\>0.151846054(20)$ \cite{LM}\\
  & & $\;\;\>0.15184(7)$ \cite{Bi}\\
5 & $\frac{25315393}{32800768}+\frac{3888425}{16400384}\frac{\sqrt{2}}{\pi}-\frac{67183425}{32800768}\frac{\arccos{(1/3)}}{\pi}$ & $\;\;\>0.07597248028...$\\ 
  & & $\;\;\>0.075972512(4)$ \cite{LM}\\
  & & $\;\;\>0.07592(6)$ \cite{Bi}\\
  & & $\;\;\>0.075978(4)$ \cite{C/M2}\\
6 & $2-\frac{81}{10}\frac{\sqrt{3}}{\pi}+\frac{38848}{1575}\frac{1}{\pi^2}$ & $\;\;\>0.03336314...$\\
  & & \\
7 & $\frac{299189248759}{290596061184}+\frac{159966456685}{435894091776}\frac{\sqrt{2}}{\pi}-\frac{292926667005}{96865353728}\frac{\arccos{(1/3)}}{\pi}$ &  $\;\;\>0.00986494662...$\\
  & &$\;\;\>0.009873(3)$ \cite{C/M2}\\
8 & $2-\frac{2511}{280}\frac{\sqrt{3}}{\pi}+\frac{17605024}{606375}\frac{1}{\pi^2}$ & $-0.00255768...$\\
  & & \\ 
9 & $\frac{2886207717678787}{2281372811001856}+\frac{2698457589952103}{5703432027504640}\frac{\sqrt{2}}{\pi}-\frac{8656066770083523}{2281372811001856}\frac{\arccos{(1/3)}}{\pi}$ & $-0.00858079817...$\\
  & & $-0.008575(3)$ \cite{C/M2}\\
10 & $2-\frac{2673}{280}\frac{\sqrt{3}}{\pi}+\frac{49048616}{1528065}\frac{1}{\pi^2}$ & $-0.01096248...$\\
  & & \\
11 &  $\frac{17357449486516274011}{11932824186709344256}+\frac{16554115383300832799}{29832060466773360640}\frac{\sqrt{2}}{\pi}-\frac{52251492946866520923}{11932824186709344256}\frac{\arccos{(1/3)}}{\pi}$ & $-0.01133719858...$\\
  & & $-0.011333(3)$ \cite{C/M2}\\
12 & $2-\frac{2187}{220}\frac{\sqrt{3}}{\pi}+\frac{11565604768}{337702365}\frac{1}{\pi^2}$ & $-0.010670281...$\\\hline\hline
\end{tabular}

\label{b4table}

\end{table}

\section{The two center formalism}
The two center formalism was invented by de Boer in 1949 \cite{dB}. This formalism is equivalent \cite{U/F} to the Mayer formalism, and in the case of hard spheres it especially useful since it allows the reduction of the dimension of the integral by $D$. The invention of this formalism is what inspired Nijboer and van Hove to confirm Boltzmann's result for $B_{4}$ in 1952 \cite{N/H}.

According to the Mayer formalism, $B_{4}$ is given by (\ref{b4mayereq1}). According to the two center formalism

\begin{eqnarray}\nonumber
\,\cs&=&-4B_{2}\,\cso(1)\\\nonumber
\,\hs&=&-\frac{4}{3}B_{2}\left(\frac{1}{2} \,
\gsqo(1)+2 \, \hhso(1)\right)\\
\,\sq&=&-\frac{8}{3}B_{2}\, \sqo(1)
\label{defvirial}
\end{eqnarray}
Thus
\begin{eqnarray}B_{4}=B_{2}\left(\frac{1}{2} \,\cso(1)+\frac{1}{2} \,
\gsqo(1)+2 \, \hhso(1)+ \, \sqo(1)\right) 
\label{g2eq}
\end{eqnarray}
Here the circles indicate points that are not integrated over, and the number 1 indicates that the distance between these two points is 1. We shall use the same notation as Nijboer and van Hove \cite{N/H}. Thus
\bea \nonumber \chi(r_{12})&:=& \,\cso\\\nonumber
(g_{1}(r_{12}))^2&:=& \,\gsqo \\\nonumber
\psi(r_{12})&:=&\, \hhso \\
\varphi(r_{12})&:=&\, \sqo 
\label{def4diag}
\eea
The functions $g_{1}(r_{12})$, $\varphi(r_{12})$ and $\psi(r_{12})$ are easily calculated for any $D$. The calculation of these diagrams in three dimensions is described in the paper by Nijboer and van Hove \cite{N/H}. It is easy to do the same calculation in higher odd dimensions, but we shall omit this since the lower order Mayer diagrams in $B_{4}$ are already known in terms of hypergeometric functions. Luban and Baram \cite{L/B} showed that 
\be \frac{\,\sq}{B_{2}^3}=\frac{2^{D+4}}{\pi}\frac{\Gamma{(D+1)}[\Gamma{(D/2+1)}]^3}{\Gamma{(3D/2+1)}[\Gamma{((D+3)/2)}]^2}\\ _3F_{2}\left(\frac{1}{2},1,\frac{-D+1}{2};\frac{D+3}{2},\frac{D+3}{2};1\right)
\label{lbring}
\ee
and 
\be \frac{\,\hs}{B_{2}^3}=-2^{D+1}D^3[\Gamma{(D/2)}]^2\int_{0}^{1}dy~ y[g_{D/2}(y)]^2
\label{lbhalf}
\ee
where
\be g_{\nu}(y)=\int_{0}^{\infty}dx~ x^{-\nu}[J_{\nu}(x)]^2J_{\nu-1}(xy)
\ee
If $D$ is odd, then according to reference \cite{GR}, p. 1071
\be _3F_{2}\left(\frac{1}{2},1,\frac{-D+1}{2};\frac{D+3}{2},\frac{D+3}{2};1\right)=\sum_{k=0}^{n}(-1)^k\frac{(2k-1)!!}{2^k}\frac{n(n-1)...(n-k+1)}{[(n+k+1)(n+k)...(n+2)]^2}
\label{fgrad}
\ee
where $D=2n+1$ and $(-1)!!=(-1)^0=1$. Joslin \cite{J} pointed out that
\begin{eqnarray}
g_{\nu}(y) = \left\lbrace \begin{array}{ll}
            \frac{2^{-\nu}y^{\nu-1}}{\Gamma{(\nu+1/2)}\Gamma{(1/2)}}\int_{2\arcsin{(y/2)}}^{\pi}d\varphi~ \cos^{2\nu}{(\varphi/2)}  & \mbox{if $y<2$} \\
            0 & \mbox{if $y\geq2$}
\end{array} \right.
\label{jg}
\end{eqnarray}
Thus, if $y<2$, $n$ is a positive integer and $D=2n+1$, then
\bea \nonumber g_{D/2}(y)=\frac{2^{-D/2}y^{D/2-1}}{\Gamma{(n+1)}\Gamma{(1/2)}}2\times~~~~~~~~~~~~~~~~~~~~~~~~~~~~~~~~~~~~~~~~~~~~~~~~~~~~~~~~~~~~~~~~~~~~~~~~~~~~~~~~~~~~~~~~~~~~~~~~~~~~\\\times\left(\frac{1}{2n+1}\frac{2^nn!}{(2n-1)!!}-\frac{y/2}{2n+1}\left\{(1-(y/2)^2)^n+\sum_{k=0}^{n-1}\frac{2^{k+1}n(n-1)...(n-k)}{(2n-1)(2n-3)...(2n-2k-1)}(1-(y/2)^2)^{n-k-1}\right\}\right)
\label{gj}
\eea
Thus using (\ref{fgrad}) and (\ref{jg}), the expressions in (\ref{lbring}) and (\ref{lbhalf}) can be explicitly computed in odd dimensions. Clearly both of these are rational numbers in odd dimensions.


\section{Integration of the complete star}

We aim to obtain a general expression for $\chi(1)$. The only dimensions lower than 12 for which the exact result has not been published before are $D=5,~7,~9,~11$. We shall calculate $\chi$ in dimensions $D=2n+1$. When $n$ is an integer, $D$ is an odd integer. However, $n$ need not be an integer. If $n$ is a half integer, then the calculation below is still valid and gives $B_{4}$ in even dimensions. If $n$ is some other positive real number, then the calculation below may be used to obtain $B_{4}$ in continous dimensions. We will use the fact that
\be r_{12}\geq 1 
\label{rrange}
\ee

According to (\ref{def4diag})
\begin{eqnarray}\chi(r_{12})=\int_{V}\int_{V} f(r_{13})f(r_{14})f(r_{23})f(r_{24})f(r_{34})d^D\mathbf{r}_{3}d^D\mathbf{r}_{4}
\end{eqnarray}
We define
\begin{eqnarray}F(h)=\int_{V} f(r_{ij})e^{2\pi i\mathbf{h}\cdot (\mathbf{r}_{i}-\mathbf{r}_{j})}d^D\mathbf{r}_{i}
\label{deff}
\end{eqnarray}
where $h=|\mathbf{h}|$. It can be shown that \cite{L/B}
\begin{eqnarray}
F(h)=-\frac{1}{h^{D/2}}J_{D/2}(2\pi h)
\label{fd}
\end{eqnarray}
where $J_{\nu}$ is a Bessel function of order $\nu$. 
We define 
\begin{eqnarray}G(\mathbf{h},r_{12})=\int_{V} f(r_{13})f(r_{23})e^{2\pi i \mathbf{h}\cdot[\mathbf{r}_{3}-\frac{1}{2}(\mathbf{r}_{1}+\mathbf{r}_{2})]}d^D\mathbf{r}_{3}
\label{defg}
\end{eqnarray}
Clearly
\begin{eqnarray}\chi(r_{12})=\int_{\mathbf{R}^D} F(h)[G(\mathbf{h},r_{12})]^2d^D\mathbf{h}
\label{intfg2}
\end{eqnarray}
In $D$ dimensions, we write $\mathbf{r}=(x_{1},x_{2},...,x_{D-1},z)=(\mathbf{x},z)$ and $\mathbf{h}=(\mathbf{h}_{x},h_{z})$. $\mathbf{r}_{1}$ and $\mathbf{r}_{2}$ are placed on the $z$ axis in such a way that $\mathbf{r}_{1}+\mathbf{r}_{2}=\mathbf{0}$. From now on, $r_{12}$ will be written as $r$. We first simplify $G(\mathbf{h},r)$. According to (\ref{defg})
\be
G(\mathbf{h},r)=2\int_{0}^{\infty}dz\cos(2\pi zh_{z})\int_{V\cap\{\mathbf{r}|z=\rm{constant}\}}d^{2n}\mathbf{x}~f([x^2+(z+r/2)^2]^{1/2})e^{2\pi i\mathbf{h}_{x}\cdot \mathbf{x}}
\label{gfinal}
\ee
where $x=|\mathbf{x}|$. The integral over the hyperplane $\{\mathbf{r}|z=\rm{constant}\}$ in (\ref{gfinal}) has the same form as the integral in (\ref{deff}) if $D$ is replaced by $2n$. It therefore follows from (\ref{fd}) that
\be G(\mathbf{h},r)=-\frac{2}{h_{x}^n}\int_{0}^{1-r/2}dz\cos{(2\pi h_{z}z)}[1-(r/2+z)^2]^{n/2}\\
J_{n}(2\pi h_{x}[1-(r/2+z)^2]^{1/2})
\label{gd}
\ee
where $h_{x}=|\mathbf{h}_{x}|$. (\ref{intfg2}) can be rewritten as
\begin{eqnarray}\chi(r)=\int_{-\infty}^{\infty}dh_{z}\int_{\mathbf{R}^{2n}} d^{2n}\mathbf{h}_{x}~F(h)[G(\mathbf{h},r)]^2
\label{intfg2new}
\end{eqnarray}
Since $F(h)$ and $G(\mathbf{h},r)$ are spherically symmetric in the hyperplane $\{\mathbf{h}|h_{z}=\rm{constant}\}$, (\ref{intfg2new}) can be simplified as
\begin{equation}
\chi(r)=\Omega_{2n-1}\int_{-\infty}^{\infty}dh_{z}\int_{0}^{\infty}dh_{x}~ h_{x}^{2n-1}F(h)[G(\mathbf{h},r)]^2
\label{iss}
\end{equation} 
where $\Omega_{2n-1}=\rm{area}(S^{2n-1})=\frac{2\pi^{n}}{\Gamma{(n)}}$. It follows from (\ref{fd}), (\ref{gd}) and (\ref{iss}) that 
\begin{eqnarray}\nonumber\chi(r)=-\frac{8\pi^n}{\Gamma{(n)}}\int_{0}^{1-r/2}dz[1-(r/2+z)^2]^{n/2}\int_{0}^{1-r/2}dz'[1-(r/2+z')^2]^{n/2}\\\nonumber
\int_{0}^{\infty}dh_{x}\frac{1}{h_{x}}J_{n}\left(2\pi[1-(r/2+z)^2]^{1/2}h_{x}\right)J_{n}\left(2\pi[1-(r/2+z')^2]^{1/2}h_{x}\right)\\
\int_{-\infty}^{\infty}dh_{z}\frac{1}{(h_{x}^2+h_{z}^2)^{D/4}}J_{D/2}\left(2\pi(h_{x}^2+h_{z}^2)^{1/2}\right)\cos{(2\pi h_{z}z)}\cos{(2\pi h_{z}z')}
\end{eqnarray}
We rewrite $\cos{(2\pi h_{z}z)}\cos{(2\pi h_{z}z')}$ as
\begin{eqnarray}\cos{(2\pi h_{z}z)}\cos{(2\pi h_{z}z')}=\frac{1}{2}\{\cos{(2\pi h_{z}(z+z'))}+\cos{(2\pi h_{z}(z-z'))}\}
\end{eqnarray}
According to reference \cite{GR}, p. 772
\begin{eqnarray}\nonumber
\int_{-\infty}^{\infty}dh_{z}\frac{1}{(h_{x}^2+h_{z}^2)^{D/4}}J_{D/2}\left(2\pi (h_{x}^2+h_{z}^2)^{1/2}\right)\cos{(2\pi h_{z}(z\pm z'))}\\
=\frac{1}{h_{x}^n}[1-(z\pm z')^2]^{n/2}J_{n}\left(2\pi h_{x}[1-(z\pm z')^2]^{1/2}\right)
\end{eqnarray}
Thus $\chi(r)$ can be reduced to a three dimensional integral. So
\begin{eqnarray}\nonumber\chi(r)=-\frac{4\pi^n}{\Gamma{(n)}}\int_{0}^{1-r/2}dz(\alpha(r/2,z)/2\pi)^n\int_{0}^{1-r/2}dz'(\alpha(r/2,z')/2\pi)^n~~~~~~~~~~~~~~~~~~~~~~~~~~~~~~~~~~~~~~~~~~~~~~~~~~~~\\
\int_{0}^{\infty}dh_{x}\frac{1}{{h_{x}}^{n+1}}J_{n}(\alpha(r/2,z) h_{x})J_{n}(\alpha(r/2,z') h_{x})\left\{(\alpha(z,z')/2\pi)^nJ_{n}(\alpha(z,z') h_{x})+(\alpha(z,-z')/2\pi)^nJ_{n}(\alpha(z,-z')h_{x})\right\}
\end{eqnarray}
where 
\begin{eqnarray}\alpha(z,z')&=&2\pi\sqrt{1-(z+z')^2}
\label{abcd}
\end{eqnarray}

Now we have to evaluate the integral $I$ given by
\begin{eqnarray}I=\int_{0}^{\infty}J_{n}(\alpha(r/2,z)x)J_{n}(\alpha(r/2,z')x)J_{n}(\alpha(z,z')x)\frac{1}{x^{n+1}}dx
\end{eqnarray}
We integrate by parts and use the recursion relations for Bessel functions
\begin{eqnarray}J_{\nu-1}(z)+J_{\nu+1}(z)=\frac{2\nu}{z}J_{\nu}(z)
\end{eqnarray}
and
\begin{eqnarray}J_{\nu-1}(z)-J_{\nu+1}(z)=2\frac{d}{dz}J_{\nu}(z)
\end{eqnarray}
Then
\begin{eqnarray}I=\frac{1}{2n}(\alpha(r/2,z)I_{\alpha(r/2,z);\alpha(r/2,z'),\alpha(z,z')}+\alpha(r/2,z') I_{\alpha(r/2,z');\alpha(r/2,z),\alpha(z,z')}+\alpha(z,z')I_{\alpha(z,z');\alpha(r/2,z),\alpha(r/2,z')})
\end{eqnarray} 
where
\begin{eqnarray}I_{\alpha;\beta,\gamma}=\int_{0}^{\infty}\frac{1}{x^n}J_{n+1}(\alpha x)J_{n}(\beta x)J_{n}(\gamma x)dx
\end{eqnarray} 
and $I_{\beta;\alpha,\gamma}$ and $I_{\gamma;\alpha,\beta}$ are defined as cyclic permutations of the same integral. We use the formula of Sonine and Dougall \cite{M/O} to calculate $I_{\alpha;\beta,\gamma}$. It says that for any positive constants $a$, $b$ and $c$
\begin{eqnarray}\nonumber\int_{0}^{\infty}J_{\mu}(at)J_{\nu}(bt)J_{\nu}(ct)t^{1-\mu}dt~~~~~~~~~~~~~~~~~~~~~~~~~~~~~~~~~~~~~~~~~~~~~~~~~~~~~\\=\frac{(bc)^{\nu}2^{-\mu+1}}{a^{\mu}\Gamma{(\mu-\nu)}\Gamma{(\nu+1/2)}\Gamma{(1/2)}}\int_{0}^{A_{a;b,c}}(a^2-b^2-c^2+2bc\cos{\varphi})^{\mu-\nu-1}\sin^{2\nu}{\varphi}~d\varphi
\end{eqnarray}
where
\bea
A_{a;b,c} = \left\lbrace \begin{array}{ll}
            0  & \mbox{if $a^2<(b-c)^2$} \\
            \arccos{\frac{b^2+c^2-a^2}{2bc}} & \mbox{if $(b-c)^2<a^2<(b+c)^2$} \\
            \pi & \mbox{if $(b+c)^2<a^2$}
\end{array} \right. 
\label{defA}
\eea
Thus
\begin{eqnarray}I_{\alpha(r/2,z);\alpha(r/2,z'),\alpha(z,z')}=\frac{2^{-n}\alpha(r/2,z')^n\alpha(z,z')^n}{\alpha(r/2,z)^{n+1}\Gamma{(n+1/2)}\Gamma{(1/2)}}\int_{0}^{A_{\alpha(r/2,z);\alpha(r/2,z'),\alpha(z,z')}}\sin^{2n}{\varphi}~d\varphi
\end{eqnarray}
We have thus reduced $\chi(r)$ to a two dimensional integral:
\begin{eqnarray}\nonumber\chi(r)=& &-\frac{2\pi^{2n}}{n\Gamma{(n+1/2)}\Gamma{(n)}\Gamma{(1/2)}}\times~~~~~~~~~~~~~~~~~~~~~~~~~~~~~~~~~~~~~~~~~~~~~~~~~~~~~~~~~~~~~~~~~~~~~~~~~~~~~~\\\nonumber
\times(& &2\int_{0}^{1-r/2}dz\int_{0}^{1-r/2}dz'(\alpha(r/2,z')/2\pi)^{2n}(\alpha(z,z')/2\pi)^{2n}\int_{0}^{A_{\alpha(r/2,z);\alpha(r/2,z'),\alpha(z,z')}}d\varphi\sin^{2n}{\varphi}\\\nonumber
&+&\int_{0}^{1-r/2}dz\int_{0}^{1-r/2}dz'(\alpha(r/2,z)/2\pi)^{2n}(\alpha(r/2,z')/2\pi)^{2n}\int_{0}^{A_{\alpha(z,z');\alpha(r/2,z),\alpha(r/2,z')}}d\varphi\sin^{2n}{\varphi}\\\nonumber
&+&2\int_{0}^{1-r/2}dz\int_{0}^{1-r/2}dz'(\alpha(r/2,z')/2\pi)^{2n}(\alpha(z,-z')/2\pi)^{2n}\int_{0}^{A_{\alpha(r/2,z);\alpha(r/2,z'),\alpha(z,-z')}}d\varphi\sin^{2n}{\varphi}\\
&+&\int_{0}^{1-r/2}dz\int_{0}^{1-r/2}dz'(\alpha(r/2,z)/2\pi)^{2n}(\alpha(r/2,z')/2\pi)^{2n}\int_{0}^{A_{\alpha(z,-z');\alpha(r/2,z),\alpha(r/2,z')}}d\varphi\sin^{2n}{\varphi})
\label{chir}
\end{eqnarray}
(The integral over $\varphi$ may be evaluated using (\ref{inteven}).) 
We need to determine which values of $z$ and $z'$ correspond to which functional form of $A_{a;b,c}$. We will use the fact that for all $z$, $z'$ for which $0\leq z,z'\leq 1-r/2$
\be \alpha(r/2,z)^2\leq(\alpha(r/2,z')+\alpha(z,z'))^2
\label{asmall}
\ee
and
\be \alpha(z,z')^2\geq(\alpha(r/2,z)-\alpha(r/2,z'))^2
\label{clarge}
\ee
Since $z'\leq 1-r/2\leq r/2$, the first inequality is obvious. The second inequality follows from the first inequality. Since $\gamma(z,-z')\geq\gamma(z,z')$ for all $z$ and $z'$, $\gamma(z,z')$ could be replaced by $\gamma(z,-z')$ in (\ref{asmall}) and (\ref{clarge}). It follows from (\ref{defA}), (\ref{asmall}) and (\ref{clarge}) that
\bea
A_{\alpha(r/2,z);\alpha(r/2,z'),\alpha(z,z')} = \left\lbrace \begin{array}{ll}
                0 & \mbox{if $\alpha(r/2,z)^2<(\alpha(r/2,z')-\alpha(z,z'))^2$} \\
                \arccos{\frac{\alpha(r/2,z')^2+\alpha(z,z')^2-\alpha(r/2,z)^2}{2\alpha(r/2,z)\alpha(z,z')}} & \mbox{if $(\alpha(r/2,z')-\alpha(z,z'))^2<\alpha(r/2,z)^2$}
\end{array}\right.
\eea
and
\bea 
A_{\alpha(z,z');\alpha(r/2,z),\alpha(r/2,z')} = \left\{ \begin{array}{ll}
                 \arccos{\frac{\alpha(r/2,z)^2+\alpha(r/2,z')^2-\alpha(z,z')^2}{2\alpha(r/2,z)\alpha(r/2,z')}}  & \mbox{if $\alpha(z,z')^2<(\alpha(r/2,z')+\alpha(r/2,z))^2$} \\
               \pi & \mbox{if $(\alpha(r/2,z')+\alpha(r/2,z))^2<\alpha(z,z')^2$}
\end{array}\right.
\eea

We need to translate the equation $\alpha(r/2,z)^2=(\alpha(r/2,z')-\alpha(z,z'))^2$ into an equation involving $z$ and $z'$. This can be done by using the definition of $\alpha$ and expanding both sides. In this way it can be shown that
\bea
A_{\alpha(r/2,z);\alpha(r/2,z'),\alpha(z,z')} = \left\{ \begin{array}{ll}
                0 & \mbox{if $z'>a_{r}(z)$} \\
                \arccos{\frac{\alpha(r/2,z')^2+\alpha(z,z')^2-\alpha(r/2,z)^2}{2\alpha(r/2,z')\alpha(z,z')}} & \mbox{if $z'<a_{r}(z)$}
                  \end{array}\right.
\eea 
and
\bea
A_{\alpha(z,z');\alpha(r/2,z),\alpha(r/2,z')} = \left\{ \begin{array}{ll}
                \pi & \mbox{if $z'>a_{r}(z)$} \\
                \arccos{\frac{\alpha(r/2,z)^2+\alpha(r/2,z')^2-\alpha(z,z')^2}{2\alpha(r/2,z)\alpha(r/2,z')}} & \mbox{if $z'<a_{r}(z)$}
                  \end{array}\right. 
\eea 
where $z'=a_{r}(z)$ is the positive root of the equation
\be3-r^2-4z^2-4zz'-4z'^2-2rz'+4r^2zz'+8rz^2z'+8rzz'^2-2rz=0
\label{eqa}
\ee
When $r=1$ this equation can be factorised as
\be (1-2z)(1-2z')(1+z+z')=0
\label{eqa1}
\ee
Hence $z'$ is undetermined whenever $z=1/2$ in this case.

It can be shown in the same way that
\bea
A_{\alpha(r/2,z);\alpha(r/2,z'),\alpha(z,-z')} = \left\{ \begin{array}{ll}
                0 & \mbox{if $z'>b_{r}(z)$} \\
                \arccos{\frac{\alpha(r/2,z')^2+{\alpha(z,-z')}^2-\alpha(r/2,z)^2}{2\alpha(r/2,z')\alpha(z,-z')}} & \mbox{if $z'<b_{r}(z)$}
                  \end{array}\right. 
\eea 
and
\bea
A_{\alpha(z,-z');\alpha(r/2,z),\alpha(r/2,z')} = \left\{ \begin{array}{ll}
                \pi & \mbox{if $z'>b_{r}(z)$} \\
                \arccos{\frac{\alpha(r/2,z)^2+\alpha(r/2,z')^2-{\alpha(z,-z')}^2}{2\alpha(r/2,z)\alpha(r/2,z')}} & \mbox{if $z'<b_{r}(z)$}
                  \end{array}
\right.
\eea
where $z'=b_{r}(z)$ is the positive root of the equation
\be2rz+2rz'-4zz'-3+4z^2+4{z'}^2+r^2=0
\label{eqb}
\ee

\begin{figure}[bt]
{\epsfig{file=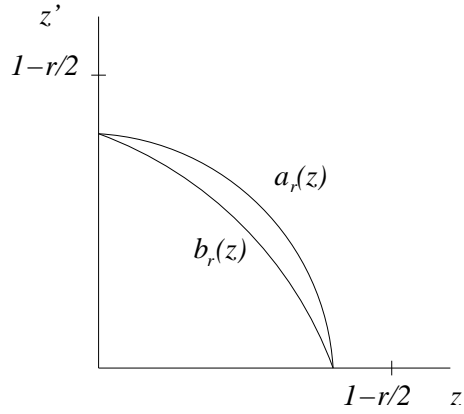, width=6cm}}
\caption{The functions $a_{r}$ and $b_{r}$. $a_{r}(z)=b_{r}(z)=0$ when $z=-\frac{r}{4}+\frac{1}{4}\sqrt{12-r^2}$.}
\label{boundaries}
\end{figure}

Since (\ref{eqa}) and (\ref{eqb}) are both symmetric in $z$ and $z'$, we could equally well write their solutions as $z=a(z')$ and $z=b(z')$ instead. $a_{r}(z)$ and $b_{r}(z)$ for $r>1$ are shown in figure \ref{boundaries}.

\begin{figure}[bt]
{\epsfig{file=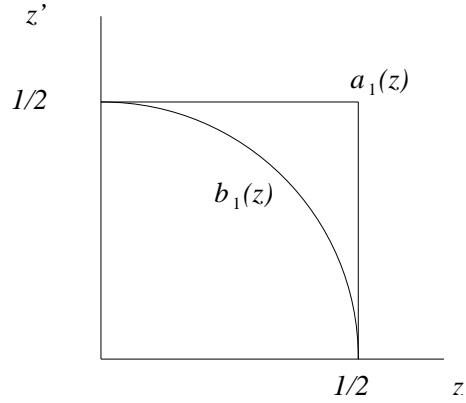, width=6cm}}
\caption{The functions $a_{1}$ and $b_{1}$ (here $r=1$). $b_{1}(z)=-\frac{1}{4}+\frac{z}{2}+\frac{3}{4}\sqrt{\frac{1}{3}(1-2z)(3+2z)}$. $b_{1}(z)=0$ when $z=\frac{1}{2}$.}
\label{boundaries1}
\end{figure}

So
\begin{eqnarray} 
\nonumber \chi(r)=&-&\frac{2\pi^{2n}}{n\Gamma{(n+1/2)}\Gamma{(n)}\Gamma{(1/2)}}\times\\\nonumber
\times(& &2\int_{0}^{1-r/2}dz\int_{0}^{a_{r}(z)}dz'(\alpha(r/2,z')/2\pi)^{2n}(\alpha(z,z')/2\pi)^{2n}\int_{0}^{\arccos{(y_{\alpha(r/2,z);\alpha(r/2,z'),\alpha(z,z')})}}d\varphi~ \sin^{2n}{\varphi}\\\nonumber
&+&\int_{0}^{1-r/2}dz\int_{0}^{a_{r}(z)}dz'(\alpha(r/2,z)/2\pi)^{2n}(\alpha(r/2,z')/2\pi)^{2n}\int_{0}^{\arccos{(y_{\alpha(z,z');\alpha(r/2,z),\alpha(r/2,z')})}}d\varphi~\sin^{2n}{\varphi}\\\nonumber 
&+&2\int_{0}^{1-r/2}dz\int_{0}^{b_{r}(z)}dz'(\alpha(r/2,z')/2\pi)^{2n}(\alpha(z,-z')/2\pi)^{2n}\int_{0}^{\arccos{(y_{\alpha(r/2,z);\alpha(r/2,z'),\alpha(z,-z')})}}d\varphi~\sin^{2n}{\varphi}\\\nonumber
&+&\int_{0}^{1-r/2}dz\int_{0}^{b_{r}(z)}dz'(\alpha(r/2,z)/2\pi)^{2n}(\alpha(r/2,z')/2\pi)^{2n}\int_{0}^{\arccos{(y_{\alpha(z,-z');\alpha(r/2,z),\alpha(r/2,z')})}}d\varphi~\sin^{2n}{\varphi}\\
&+&\int_{0}^{1-r/2}dz\int_{b_{r}(z)}^{1/2}dz'(\alpha(r/2,z)/2\pi)^{2n}(\alpha(r/2,z')/2\pi)^{2n}\int_{0}^{\pi}d\varphi~\sin^{2n}{\varphi})
\eea
where $y_{\alpha;\beta,\gamma}=\frac{\beta^2+\gamma^2-\alpha^2}{2\beta\gamma}$. As $r$ tends to 1, it follows from (\ref{eqa1}) that $a_{r}(z)$ takes the value 1/2 for all $z$, as shown in figure \ref{boundaries1}. In the special case $r=1$ the integral simplifies to
\begin{eqnarray} 
\nonumber \chi(1)=&-&\frac{2\pi^{2n}}{n\Gamma{(n+1/2)}\Gamma{(n)}\Gamma{(1/2)}}\times\\\nonumber
\times(& &2\int_{0}^{1/2}dz\int_{0}^{1/2}dz'(\alpha(1/2,z')/2\pi)^{2n}(\alpha(z,z')/2\pi)^{2n}\int_{0}^{\arccos{(y_{\alpha(1/2,z);\alpha(1/2,z'),\alpha(z,z')})}}d\varphi~\sin^{2n}{\varphi}\\\nonumber
&+&\int_{0}^{1/2}dz\int_{0}^{1/2}dz'(\alpha(1/2,z)/2\pi)^{2n}(\alpha(1/2,z')/2\pi)^{2n}\int_{0}^{\arccos{(y_{\alpha(z,z');\alpha(1/2,z),\alpha(1/2,z')})}}d\varphi~\sin^{2n}{\varphi}\\\nonumber 
&+&2\int_{0}^{1/2}dz\int_{0}^{b_{1}(z)}dz'(\alpha(1/2,z')/2\pi)^{2n}(\alpha(z,-z')/2\pi)^{2n}\int_{0}^{\arccos{(y_{\alpha(1/2,z);\alpha(1/2,z'),\alpha(z,-z')})}}d\varphi~\sin^{2n}{\varphi}\\\nonumber
&+&\int_{0}^{1/2}dz\int_{0}^{b_{1}(z)}dz'(\alpha(1/2,z)/2\pi)^{2n}(\alpha(1/2,z')/2\pi)^{2n}\int_{0}^{\arccos{(y_{\alpha(z,-z');\alpha(1/2,z),\alpha(1/2,z')})}}d\varphi~\sin^{2n}{\varphi}\\
&+&\int_{0}^{1/2}dz\int_{b_{1}(z)}^{1/2}dz'(\alpha(1/2,z)/2\pi)^{2n}(\alpha(1/2,z')/2\pi)^{2n}\int_{0}^{\pi}d\varphi~\sin^{2n}{\varphi})
\eea
After integration by parts, this gives integrals of the type 
\begin{eqnarray}\int\frac{p(x)}{q(x)\sqrt{a+bx+cx^2}}dx
\end{eqnarray}
where $p$ and $q$ are polynomials. Using Maple it was thus possible to calculate $\chi(1)$ for $D=5,~7,~9,~11$. We may now obtain $\,\cs$ from (\ref{defvirial}). Since $\,\sq$ and $\,\hs$ can be obtained from (\ref{lbring}) and (\ref{lbhalf}), we have found $B_{4}$. We use the more compact Ree Hoover $\tilde{f}$ formalism \cite{ReeHoover} to present the results. In this formalism $B_{4}$ consists of only two diagrams instead of three. Here 
\be
\tilde{f}(r_{ij})-f(r_{ij})=1
\ee
Thus
\bea \nonumber\emptyset&=&\,\cs\\
\,\rhring&=&\,\cs+2\,\hs+\,\sq
\eea
and
\bea B_{4}=\frac{1}{4}\emptyset-\frac{3}{8}\,\rhring
\eea
The final answer is given in tables \ref{b4table}, \ref{cstable} and \ref{ringtable}. The numerical values of references \cite{C/M2} and \cite{Bi} agree with the exact result.


\begin{table}[h]
\caption{Exact and numerical \cite{C/M2} values of the Ree Hoover complete star}

\begin{tabular}{|l|l|l|} \hline\hline
$D$ &
$\frac{\emptyset}{4B_{2}^3}$ & decimal expansion\\ \hline
3 & $-\frac{89}{280}-\frac{219}{1120}\frac{\sqrt{2}}{\pi}+\frac{4131}{2240}\frac{\arccos{(1/3)}}{\pi}$ & 0.31672598803...\\

 & & 0.31673(2)\\
5 & $-\frac{163547}{128128}-\frac{3888425}{8200192}\frac{\sqrt{2}}{\pi}+\frac{67183425}{16400384}\frac{\arccos{(1/3)}}{\pi}$ & 0.11520591833...\\

 & & 0.115211(3)\\
7 & $-\frac{283003297}{141892608}-\frac{159966456685}{217947045888}\frac{\sqrt{2}}{\pi}+\frac{292926667005}{48432676864}\frac{\arccos{(1/3)}}{\pi}$ & 0.04492254969...\\

 & & 0.044927(2)\\ 
9 & $-\frac{88041062201}{34810986496}-\frac{2698457589952103}{2851716013752320}\frac{\sqrt{2}}{\pi}+\frac{8656066770083523}{1140686405500928}\frac{\arccos{(1/3)}}{\pi}$ & 0.01828214224...\\

 & & 0.018286(1)\\
11 & $-\frac{66555106087399}{22760055898112}-\frac{16554115383300832799}{14916030233386680320}\frac{\sqrt{2}}{\pi}+\frac{52251492946866520923}{5966412093354672128}\frac{\arccos{(1/3)}}{\pi}$ & 0.00766164876... \\

 & & 0.0076638(8)\\\hline\hline
\end{tabular}
\label{cstable}
\end{table}

\begin{table}[h]
\caption{Exact and numerical \cite{C/M2} values of the Ree Hoover ring}

\begin{tabular}{|l|l|l|} \hline\hline
$D$ &
$-\frac{3\,\rhring}{8B_{2}^3}$ & decimal expansion\\ \hline

3 & $\frac{4131}{4480}+\frac{657}{2240}\frac{\sqrt{2}}{\pi}-\frac{12393}{4480}\frac{\arccos{(1/3)}}{\pi}$ & $-0.02977648205...$ \\

 & & $-0.029781(8)$\\

5 & $\frac{67183425}{32800768}+\frac{11665275}{16400384}\frac{\sqrt{2}}{\pi}-\frac{201550275}{32800768}\frac{\arccos{(1/3)}}{\pi}$ & $-0.03923343804...$\\

 & & $-0.039233(3)$\\
7 & $\frac{292926667005}{96865353728}+\frac{159966456685}{145298030592}\frac{\sqrt{2}}{\pi}-\frac{878780001015}{96865353728}\frac{\arccos{(1/3)}}{\pi}$ & $-0.03505760307...$  \\

 & & $-0.035055(3)$\\ 
9 & $\frac{8656066770083523}{2281372811001856}+\frac{8095372769856309}{5703432027504640}\frac{\sqrt{2}}{\pi}-\frac{25968200310250569}{2281372811001856}\frac{\arccos{(1/3)}}{\pi}$ & $-0.02686294042...$ \\

 & & $-0.026861(3)$\\
11 &  $\frac{52251492946866520923}{11932824186709344256}+\frac{49662346149902498397}{29832060466773360640}\frac{\sqrt{2}}{\pi}-\frac{156754478840599562769}{11932824186709344256}\frac{\arccos{(1/3)}}{\pi}$ & $-0.01899884734...$ \\

 & & $-0.018997(3)$ \\\hline\hline
\end{tabular}
\label{ringtable}
\end{table}

\section{Discussion}

\begin{table}[h]
\caption{Contributions to the fifth virial coefficient in three dimensions \cite{KimH}, \cite{RH}}

\begin{tabular}{|l|l|} \hline\hline
diagram & exact value  \\ \hline
$E5/B_{2}^4$ & $ -\frac{40949}{10752}$ \\
$E6\alpha/B_{2}^4$ & $ \;\;\>\frac{68419}{26880}$  \\
$E6\beta/B_{2}^4$ & $\;\;\>\frac{82}{35}$ \\
$E7\alpha/B_{2}^4$ & $-\frac{34133}{17920}$  \\
$E7\beta/B_{2}^4$ & $-\frac{18583}{5376}+\frac{33291}{9800}\frac{\sqrt{3}}{\pi}$ \\
$E7\gamma/B_{2}^4$ & $-\frac{73491}{35840}$ \\
$E8\alpha/B_{2}^4$ & unknown \\
$E8\beta/B_{2}^4$ & $-\frac{35731}{6720}+\frac{1458339}{627200}\frac{\sqrt{2}}{\pi}-\frac{33291}{9800}\frac{\sqrt{3}}{\pi}+\frac{683559}{35840}\frac{\arccos{(1/3)}}{\pi}$ \\ 
$E9/B_{2}^4$ & unknown \\
$E10/B_{2}^4$ & unknown \\ \hline\hline
\end{tabular}
\label{b5table}
\end{table}

Table \ref{b4table} shows exact and numerical values of the fourth virial coefficient. Typically the relative error of the numerical value is of order $10^{-4}$. Recently Clisby and McCoy \cite{C/M2} calculated higher order coefficients using Monte Carlo methods. It is seen that the relative error increases with the order of the coefficient, which is one of the reasons why it is desirable to find analytic values. Table \ref{b5table} shows the known exact values of diagrams of the fifth virial coefficient in 3 dimensions. The diagrams $E7\beta$ and $E8\beta$ have the same coefficient, so there is so far no total contribution of $\sqrt{3}/\pi$.
We make the following conjecture:

\begin{conjecture}
In any dimension, the hard sphere potential (\ref{hspot}) allows the analytic computation of every virial coefficient $B_{n}$.
\end{conjecture}
No one has so far been able to prove this conjecture, but since the complete star of any number of points can be expressed in terms of an integral involving the same functions $F$ and $G$ used in section III, there is good reason to assume that it is true and can be proven.

\begin{acknowledgments}
The author wishes to thank Dr. Nathan Clisby for numerical integration and Prof. Barry McCoy for useful discussions. This work has been supported by NSF grant 0302758.
\end{acknowledgments}

\end{document}